\documentclass[
 aip,
 jcp,
 10pt,
 reprint,
 floatfix,
 numbers
]{revtex4-1}

\usepackage{amsmath,amssymb}
\usepackage{graphicx}
\usepackage{dcolumn}
\usepackage{bm}
\usepackage{float}
\usepackage[utf8]{inputenc}
\usepackage[T1]{fontenc}
\usepackage{mathptmx}
\usepackage{etoolbox}
\usepackage{booktabs}
\usepackage{url}

\setcitestyle{numbers,square}

\makeatletter
\def\@email#1#2{%
 \endgroup
 \patchcmd{\titleblock@produce}
  {\frontmatter@RRAPformat}
  {\frontmatter@RRAPformat{\produce@RRAP{*#1\href{mailto:#2}{#2}}}\frontmatter@RRAPformat}
  {}{}
}%
\makeatother

\begin{document}

\preprint{AIP/123-QED}

\title{Learnable Viscosity Modulation in Physics-Informed Neural Networks for Incompressible Flow Reconstruction}

\author{Ke Xu}
\altaffiliation{These authors contributed equally to this work.}
\affiliation{Nanophotonics and Biophotonics Key Laboratory of Jilin Province, School of Physics, Changchun University of Science and Technology, Changchun 130022, P.\ R.\ China}

\author{Ze Tao}
\altaffiliation{These authors contributed equally to this work.}
\affiliation{Nanophotonics and Biophotonics Key Laboratory of Jilin Province, School of Physics, Changchun University of Science and Technology, Changchun 130022, P.\ R.\ China}

\author{Fujun Liu}
\email{fjliu@cust.edu.cn}
\affiliation{Nanophotonics and Biophotonics Key Laboratory of Jilin Province, School of Physics, Changchun University of Science and Technology, Changchun 130022, P.\ R.\ China}

\date{\today}

\begin{abstract}
Accurately and stably solving the incompressible Navier--Stokes equations with physics-informed neural networks (PINNs) remains challenging, particularly for sparse or noisy observations and for flow regimes in which the local balance among convection, diffusion, and pressure is difficult to capture. To address this issue, we propose a framework, denoted as LVM-PINN, which incorporates a learnable viscosity modulation (LVM) mechanism into the PINN residual. Specifically, the model predicts a spatiotemporal scalar field that is embedded directly into the viscous diffusion term of the momentum equations, thereby enabling adaptive modulation of the local dissipation strength during training. This modification improves optimization stability while enhancing the representation of complex flow structures. The effect of the proposed mechanism is further examined through a controlled ablation setting with an otherwise unchanged network architecture, as well as through comparisons with GRU- and residual-attention-based backbone baselines. Numerical experiments on two-dimensional benchmark problems, including the Kovasznay flow and two manufactured forcing flows, show that the proposed framework yields more stable training behavior and more accurate flow reconstruction under sparse and noisy data conditions.
\end{abstract}
\keywords{Physics-informed neural networks, incompressible Navier--Stokes equations, viscosity modulation, flow reconstruction, sparse observations}

\maketitle

\maketitle

\section{Introduction}

Accurate reconstruction of incompressible flow fields from sparse and noisy observations remains a fundamental challenge in computational fluid mechanics. While classical numerical solvers produce high-fidelity solutions when governing equations, boundary conditions, and discretizations are fully specified, their computational cost becomes substantial in settings involving repeated inversions, data assimilation, or limited observations. Conversely, purely data-driven neural network models offer greater efficiency post-training but often suffer from poor physical consistency and limited generalization under sparse or noisy data conditions. These limitations have motivated the development of hybrid learning frameworks that explicitly combine observational data with the governing physical laws of the flow. Physics-informed neural networks (PINNs) provide a robust solution by embedding these governing equations directly into the loss function, ensuring the model satisfies both empirical constraints and physical laws \cite{Raissi2019,xing2025modeling,raissi2020hidden}. Since their introduction, PINNs have been successfully applied to a wide range of forward and inverse problems involving nonlinear partial differential equations (PDEs) across fluid mechanics and multiphysics systems \cite{Raissi2019,xing2025modeling,tao2025analytical,jin2021nsf,mao2020highspeed,meng2019ppinn}. Despite these successes, standard PINN formulations often encounter optimization difficulties in complex flow regimes\cite{xing2025modeling,krishnapriyan2021failure}. This instability is particularly evident when observations are sparse, unevenly distributed, or noisy, making it difficult for a fixed residual representation to capture the delicate local balances among convection, diffusion, and pressure \cite{Raissi2019,tao2025lnn,wang2022understanding,bischof2021multiobjective,wu2022adaptive}.

To address this inherent limitation, we propose a novel framework, denoted as LVM-PINN, which incorporates a learnable viscosity modulation (LVM) mechanism into the standard PINN architecture. In addition to predicting the primary fluid variables, the neural network outputs an auxiliary spatiotemporal scalar field, denoted by $Nu_\theta(x,y,t)$. This field is dynamically embedded into the diffusion term of the momentum residual as an effective viscosity correction. Consequently, the local diffusion strength can be adaptively adjusted during training while strictly preserving the original mathematical structure of the incompressible Navier-Stokes equations. From a physical modeling perspective, this approach conceptually parallels the broader idea of effective or eddy viscosity used in fluid mechanics to modify momentum transport or diffusion strength \cite{Pope2000}. However, within our framework, the modulation field is not treated as an independently observed physical quantity. Instead, it serves purely as a learnable mathematical correction within the residual operator to improve optimization stability and enhance overall reconstruction quality. To rigorously isolate and quantify the effect of this mechanism, we evaluate two controlled configurations within the identical PINN architecture. The learn configuration actively embeds the modulation field into the diffusion term of the governing equation residual. In contrast, the off configuration completely excludes this modulation field from the residual construction while keeping the network architecture and output dimensions entirely unchanged. This controlled ablation design allows us to explicitly assess the contribution of the proposed physical mechanism without conflating it with variations in network size or optimization strategy.

To evaluate the efficacy of this modulated architecture, our numerical study investigates three distinct benchmark problems. The first is the classical Kovasznay flow, which serves as an unforced analytical benchmark characterized by nontrivial nonlinear coupling. The subsequent two cases are manufactured forcing flows, where analytical fields are explicitly prescribed and the corresponding forcing terms are constructed consistently from the governing momentum equations. Together, these three benchmarks encompass varying levels of flow complexity, providing a comprehensive and unified basis for testing the proposed model under sparse and noisy observational conditions. Beyond the internal ablation comparison between the learn and off configurations, the framework is also evaluated against two alternative backbone baselines, namely a Gated Recurrent Unit (GRU)-based PINN and a residual-attention (ResAttn) PINN. This expanded experimental design ensures that the proposed framework is rigorously examined against both its own internal mechanisms and representative architectural alternatives.

Based on this methodology, the main contributions of this work are threefold. First, we propose the LVM-PINN framework, a novel extension of standard PINNs that embeds a learnable viscosity modulation (LVM) field directly into the diffusion term of the momentum residual. Second, we construct a carefully controlled comparison between active and inactive modulation configurations to unambiguously isolate the performance enhancements derived from the mechanism itself. Third, we perform a systematic empirical evaluation against both internal ablations and external neural network baselines on analytical and manufactured incompressible flow benchmarks. These evaluations collectively demonstrate that the proposed LVM-PINN framework yields significantly improved optimization stability and highly competitive reconstruction accuracy across diverse flow regimes. The remainder of this paper is organized to detail these findings. Section 2 presents the rigorous mathematical formulation of the proposed LVM-PINN framework. Section 3 introduces the benchmark problems and outlines the specific experimental settings. Section 4 summarizes the detailed implementation procedures. Section 5 presents the numerical results alongside a comprehensive comparative analysis. Finally, Section 6 provides concluding remarks.

\section{Mathematical Formulation}

PINNs recast the solution of partial differential equations into a constrained neural network training problem \cite{Raissi2019}. Rather than relying solely on observational data, a PINN approximates the target fluid fields using a neural network and optimizes its parameters by minimizing a composite loss function that enforces both empirical data consistency and physical conservation laws. In the present study, the primary objective is to reconstruct two-dimensional incompressible flow fields from sparse and noisy observations while strictly preserving compliance with the governing Navier-Stokes equations. Let $\mathbf{x}=(x,y,t)$ denote the spatiotemporal coordinates, $\mathbf{u}=(u,v)$ denote the velocity field, and $p$ denote the pressure field. In nondimensional form, the governing momentum and continuity equations are written as
\begin{equation}
u_t + u u_x + v u_y + p_x - \frac{1}{Re}(u_{xx}+u_{yy}) = f_x,
\label{eq:ns_u}
\end{equation}
\begin{equation}
v_t + u v_x + v v_y + p_y - \frac{1}{Re}(v_{xx}+v_{yy}) = f_y,
\label{eq:ns_v}
\end{equation}
\begin{equation}
u_x + v_y = 0,
\label{eq:ns_c}
\end{equation}
where $Re$ is the Reynolds number, and $f_x$ and $f_y$ denote the prescribed forcing terms. For unforced analytical benchmarks, these forcing terms naturally vanish such that $f_x=f_y=0$, whereas for manufactured forcing cases, they are explicitly prescribed so that a selected analytical solution exactly satisfies the governing equations. In a standard PINN setting, the neural network predicts these physical variables, and the partial differential equation residuals are evaluated via automatic differentiation to penalize non-physical predictions during the optimization process \cite{Raissi2019}.

Building upon this foundational framework, the present work introduces an additional learnable scalar field designed to dynamically modulate the viscous diffusion term. Specifically, the neural network, parameterized by weights and biases $\theta$, is defined as
\begin{equation}
(u_\theta, v_\theta, p_\theta, Nu_\theta)=\mathcal{N}_\theta(x,y,t),
\label{eq:network_output}
\end{equation}
where $u_\theta$ and $v_\theta$ are the predicted velocity components, $p_\theta$ is the predicted pressure field, and $Nu_\theta$ is an auxiliary learnable modulation field. This additional variable is strategically introduced to adaptively adjust the local diffusion strength during the training phase. To implement this mechanism, an effective viscosity is defined as
\begin{equation}
\nu_{\mathrm{eff}}=\nu\bigl(1+Nu_\theta(x,y,t)\bigr),
\qquad
\nu=\frac{1}{Re},
\label{eq:nueff}
\end{equation}
so that the diffusion coefficient is no longer constrained uniformly across the computational domain but is allowed to vary according to the local flow state inferred by the network. From a fluid modeling perspective, this mathematical construction relates to the broader concept of effective or eddy viscosity, where an additional viscosity-like quantity modifies momentum transport or overall diffusion strength \cite{Pope2000}. Within the proposed LVM-PINN framework, however, $Nu_\theta$ is neither introduced as an independently observed physical variable nor treated as a conventional turbulence model parameter. Instead, it serves purely as a learnable viscosity modulation field embedded directly into the PINN residual, fulfilling the specific purpose of adaptively regulating the local dissipation level to stabilize training.

By integrating this effective viscosity, the residuals of the momentum and continuity equations in the active learn configuration are formulated as
\begin{equation}
\begin{aligned}
r_x^{\mathrm{learn}} ={}&
u_{\theta,t}+u_\theta u_{\theta,x}+v_\theta u_{\theta,y}+p_{\theta,x} \\
&-\frac{1}{Re}(1+Nu_\theta)(u_{\theta,xx}+u_{\theta,yy})-f_x ,
\end{aligned}
\label{eq:rx_learn}
\end{equation}
\begin{equation}
\begin{aligned}
r_y^{\mathrm{learn}} ={}&
v_{\theta,t}+u_\theta v_{\theta,x}+v_\theta v_{\theta,y}+p_{\theta,y} \\
&-\frac{1}{Re}(1+Nu_\theta)(v_{\theta,xx}+v_{\theta,yy})-f_y ,
\end{aligned}
\label{eq:ry_learn}
\end{equation}
\begin{equation}
r_c^{\mathrm{learn}} = u_{\theta,x}+v_{\theta,y}.
\label{eq:rc_learn}
\end{equation}
To rigorously isolate and evaluate the contribution of this modulation mechanism, a controlled ablation setting, denoted as the off configuration, is simultaneously considered. In this baseline case, the neural network architecture and output dimensions remain identical, but the predicted modulation field $Nu_\theta$ is strictly excluded from the governing equation residuals. Consequently, the corresponding residuals revert to the standard formulation:
\begin{equation}
\begin{aligned}
r_x^{\mathrm{off}} ={}&
u_{\theta,t}+u_\theta u_{\theta,x}+v_\theta u_{\theta,y}+p_{\theta,x} \\
&-\frac{1}{Re}(u_{\theta,xx}+u_{\theta,yy})-f_x ,
\end{aligned}
\label{eq:rx_off}
\end{equation}
\begin{equation}
\begin{aligned}
r_y^{\mathrm{off}} ={}&
v_{\theta,t}+u_\theta v_{\theta,x}+v_\theta v_{\theta,y}+p_{\theta,y} \\
&-\frac{1}{Re}(v_{\theta,xx}+v_{\theta,yy})-f_y ,
\end{aligned}
\label{eq:ry_off}
\end{equation}
\begin{equation}
r_c^{\mathrm{off}} = u_{\theta,x}+v_{\theta,y}.
\label{eq:rc_off}
\end{equation}
This controlled mathematical construction ensures that any performance differences between the learn and off configurations can be attributed solely to the learnable viscosity modulation, completely preventing the conflation of results with variations in network capacity or underlying optimization strategies.

The overall training process is driven by minimizing a composite objective function comprising a data-fitting term and a physics-residual term. Given a set of sparse observations denoted by $\mathcal{D}_d=\{(x_i,y_i,t_i,u_i,v_i,p_i)\}_{i=1}^{N_d}$, the empirical data loss is defined as
\begin{equation}
\begin{aligned}
\mathcal{L}_{\mathrm{data}} ={}&
\frac{1}{N_d}\sum_{i=1}^{N_d}
\Bigl[
|u_\theta(\mathbf{x}_i)-u_i|^2 \\
&\qquad + |v_\theta(\mathbf{x}_i)-v_i|^2
+ |p_\theta(\mathbf{x}_i)-p_i|^2
\Bigr] .
\end{aligned}
\label{eq:ldata}
\end{equation}
Simultaneously, a set of collocation points $\mathcal{D}_r=\{(x_j,y_j,t_j)\}_{j=1}^{N_r}$ is sampled throughout the computational domain to enforce the physical constraints, yielding the physics loss
\begin{equation}
\mathcal{L}_{\mathrm{eq}}
=
\frac{1}{N_r}\sum_{j=1}^{N_r}
\Bigl[
r_x(\mathbf{x}_j)^2+r_y(\mathbf{x}_j)^2+r_c(\mathbf{x}_j)^2
\Bigr],
\label{eq:leq}
\end{equation}
where the specific residuals are selected according to either the learn or the off configuration. For the standard unforced benchmark settings, the total loss is formulated as a weighted sum,
\begin{equation}
\mathcal{L}
=
\mathcal{L}_{\mathrm{data}}
+
\lambda_{\mathrm{eq}}\mathcal{L}_{\mathrm{eq}},
\label{eq:loss_total}
\end{equation}
where $\lambda_{\mathrm{eq}}$ dictates the relative importance of the physics residual. For the manufactured forcing cases, a reference modulation field $Nu^\ast(x,y,t)$ is mathematically accessible, allowing for the introduction of an auxiliary supervision term defined as
\begin{equation}
\mathcal{L}_{Nu}
=
\frac{1}{N_r}\sum_{j=1}^{N_r}
\left|Nu_\theta(\mathbf{x}_j)-Nu^\ast(\mathbf{x}_j)\right|^2.
\label{eq:lnu}
\end{equation}
This auxiliary term is subsequently integrated into an extended total loss function written as
\begin{equation}
\mathcal{L}
=
\mathcal{L}_{\mathrm{data}}
+
\lambda_{\mathrm{eq}}\mathcal{L}_{\mathrm{eq}}
+
\lambda_{Nu}\mathcal{L}_{Nu}.
\label{eq:loss_forced}
\end{equation}
Finally, the optimal network parameters are determined by solving the overarching minimization problem
\begin{equation}
\theta^\ast=\arg\min_\theta \mathcal{L}(\theta),
\label{eq:opt_problem}
\end{equation}
which is computationally resolved via gradient-based optimization utilizing automatic differentiation. Through this comprehensive mathematical formulation, the proposed LVM-PINN framework structurally extends the standard PINN approach by incorporating adaptive viscosity modulation while maintaining a unified, end-to-end differentiable training procedure.

\section{Benchmark Problems and Experimental Settings}

To rigorously evaluate the proposed LVM-PINN framework across distinct flow regimes, we consider three comprehensive benchmark problems encompassing the classical Kovasznay flow and two manufactured forcing scenarios. Following the operator-based formulation established in this work, all benchmark flow fields are governed by a unified mathematical representation, expressed as
\begin{equation}
\mathcal{R}_m(\mathbf{u},p)=\mathbf{f},
\label{eq:benchmark_rm}
\end{equation}
\begin{equation}
\mathcal{R}_c(\mathbf{u})=0,
\label{eq:benchmark_rc}
\end{equation}
where $\mathcal{R}_m$ denotes the momentum residual operator and $\mathcal{R}_c$ represents the incompressibility operator. For all subsequent experiments, high-fidelity reference data are initially generated on structured spatiotemporal grids and subsequently subsampled randomly to simulate sparse observational conditions, with prescribed Gaussian noise added to rigorously test algorithmic robustness.

\subsection{Kovasznay Flow}

The classical Kovasznay flow serves as our primary analytical benchmark, selected specifically to assess the model's capacity to accurately recover exponentially decaying vortex structures driven by nonlinear convection. The computational domain for this steady, unforced flow is defined as
\begin{equation}
x \in [-0.5,1.0], \qquad y \in [-0.5,1.5], \qquad t \in [0,1],
\end{equation}
and the Reynolds number is fixed at
\begin{equation}
Re=100.
\end{equation}
By defining the spatial decay rate as
\begin{equation}
\lambda=\frac{Re}{2}-\sqrt{\frac{Re^2}{4}+4\pi^2},
\end{equation}
the exact analytical solution for the velocity and pressure fields is given by
\begin{equation}
u(x,y)=1-e^{\lambda x}\cos(2\pi y),
\end{equation}
\begin{equation}
v(x,y)=\frac{\lambda}{2\pi}e^{\lambda x}\sin(2\pi y),
\end{equation}
\begin{equation}
p(x,y)=\frac{1}{2}\bigl(1-e^{2\lambda x}\bigr).
\end{equation}
Because this specific benchmark represents a naturally occurring unforced flow equilibrium, the forcing vector is strictly zero, such that
\begin{equation}
\mathbf{f}=\mathbf{0}.
\end{equation}
To establish a highly sparse training environment for the numerical experiments, only $5\%$ of the total available observations are retained, and these sampled data points are further corrupted by Gaussian noise with an amplitude of $0.01$ to challenge the neural network's physical consistency.

\subsection{Manufactured Forcing Flow I}

Building upon the unforced Kovasznay benchmark, we introduce Manufactured Forcing Flow I to evaluate the model's performance under explicit external forcing conditions and elevated inertial effects. The computational domain for this highly dynamic scenario is defined as
\begin{equation}
x \in [-1.2,1.1], \qquad y \in [-1.0,1.0], \qquad t \in [0,1],
\end{equation}
with the Reynolds number significantly increased to
\begin{equation}
Re=2500.
\end{equation}
Unlike the naturally occurring Kovasznay flow, this case utilizes the method of manufactured solutions to rigidly control the flow topology rather than treating it as an unforced exact Navier-Stokes solution. Specifically, the analytical fields $(\mathbf{u},p,Nu^\ast)$ are explicitly prescribed a priori, and the corresponding forcing term is analytically derived through the momentum residual operator defined in Eq.~\eqref{eq:benchmark_rm}. To test the robustness of the learning mechanism under these forced conditions, the training dataset is heavily restricted to merely $6\%$ of the global observations, accompanied by Gaussian noise with an amplitude of $0.008$. The explicit analytical expressions detailing these manufactured fields and the associated reference modulation field are provided in Appendix~\ref{app:manufactured_cases} to ensure complete reproducibility.

\subsection{Manufactured Forcing Flow II}

To further validate the framework across varying spatial gradients, Manufactured Forcing Flow II is designed to exhibit smoother spatiotemporal flow structures while retaining nontrivial, forcing-driven internal dynamics. The computational domain for this final benchmark is specified as
\begin{equation}
x \in [-1.0,1.0], \qquad y \in [-1.2,0.8], \qquad t \in [0,1],
\end{equation}
and the Reynolds number is further elevated to
\begin{equation}
Re=2800.
\end{equation}
Following the precise methodology established in the first manufactured case, the analytical fields $(\mathbf{u},p,Nu^\ast)$ are prescribed initially, and the necessary forcing components are subsequently computed via the exact operator relation $\mathbf{f}=\mathcal{R}_m(\mathbf{u},p)$. For the corresponding numerical experiments, $5.5\%$ of the spatial observations are randomly retained to form the training set, which is similarly subjected to Gaussian noise with an amplitude of $0.008$. Through this systematic progression from unforced equilibrium flows to complex manufactured dynamics, we establish a robust testing paradigm, with the full explicit analytical expressions for this final smooth forcing benchmark also detailed in Appendix~\ref{app:manufactured_cases}.

\section{Implementation Details}

All models are implemented within a unified physics-informed learning framework. Spatial and temporal derivatives required for the governing equations are computed analytically utilizing automatic differentiation, while the network parameters are iteratively optimized via gradient-based training algorithms. For clarity, Nu\_learn denotes the proposed model with active viscosity modulation, whereas Nu\_off denotes its controlled ablation counterpart with the modulation field excluded from the momentum residual. GRU and ResAttn denote the two alternative backbone baselines, corresponding to a GRU-based PINN and a residual-attention PINN, respectively. To establish the training environment for each benchmark, high-fidelity reference data are first generated from the corresponding analytical or manufactured solutions. These full-field continuous data are subsequently subsampled according to explicitly prescribed observation ratios and noise levels to simulate realistic, sparse, and noisy measurement conditions. Concurrently, an independent set of collocation points is densely sampled throughout the computational domain to continuously evaluate the partial differential equation residuals and enforce physical constraints during the optimization process.

Utilizing these prepared datasets, the proposed LVM-PINN framework is rigorously evaluated in two distinct internal configurations, denoted as Nu\_learn and Nu\_off. In the active Nu\_learn configuration, the learnable modulation field $Nu_\theta(x,y,t)$ is dynamically embedded into the diffusion term of the momentum residual through the defined effective viscosity $\nu_{\mathrm{eff}}=\nu(1+Nu_\theta)$. Conversely, in the Nu\_off ablation setting, the overall network capacity and architecture remain strictly unchanged, but the predicted modulation field is entirely excluded from the governing equation residuals. To contextualize these internal ablation results, two alternative neural network backbones, specifically a GRU-based PINN and a residual-attention PINN (ResAttn), are also evaluated under identical benchmark conditions. This carefully structured experimental design ensures that the specific performance enhancements derived from the proposed viscosity modulation mechanism can be explicitly distinguished from variations caused merely by altering the underlying neural architecture.

Throughout the training process, model convergence is systematically monitored by independently tracking the total loss, the empirical data loss, and the physics equation loss. For the manufactured forcing cases, an auxiliary supervision term associated with the explicit reference modulation field $Nu^\ast$ is additionally incorporated into the objective function when applicable, as formally defined in Eq.~\eqref{eq:loss_forced}. Upon completion of the training phase, a comprehensive final evaluation is conducted on the reconstructed pressure field and both velocity components. This terminal assessment relies on rigorous quantitative error statistics combined with qualitative spatial field visualizations, thereby providing a consistent and robust measure of training stability, physical consistency, and overall flow reconstruction quality across all compared methods.

\section{Results and Analysis}

This section details the comparative performance of the proposed LVM-PINN framework across three distinct hydrodynamic regimes: the classical Kovasznay flow, manufactured forcing flow I, and manufactured forcing flow II. For each benchmark scenario, the active modulation model, Nu\_learn, is systematically evaluated against its own ablation counterpart, Nu\_off, as well as against two established alternative neural network baselines, specifically a residual-attention architecture (ResAttn) and a GRU-based PINN. The ensuing discussion synthesizes the training loss histories, the qualitative reconstructed flow fields, and the quantitative error statistics compiled in the corresponding tables to provide a comprehensive assessment of model performance.

\begin{figure}[H]
    \centering
    \includegraphics[width=0.80\columnwidth]{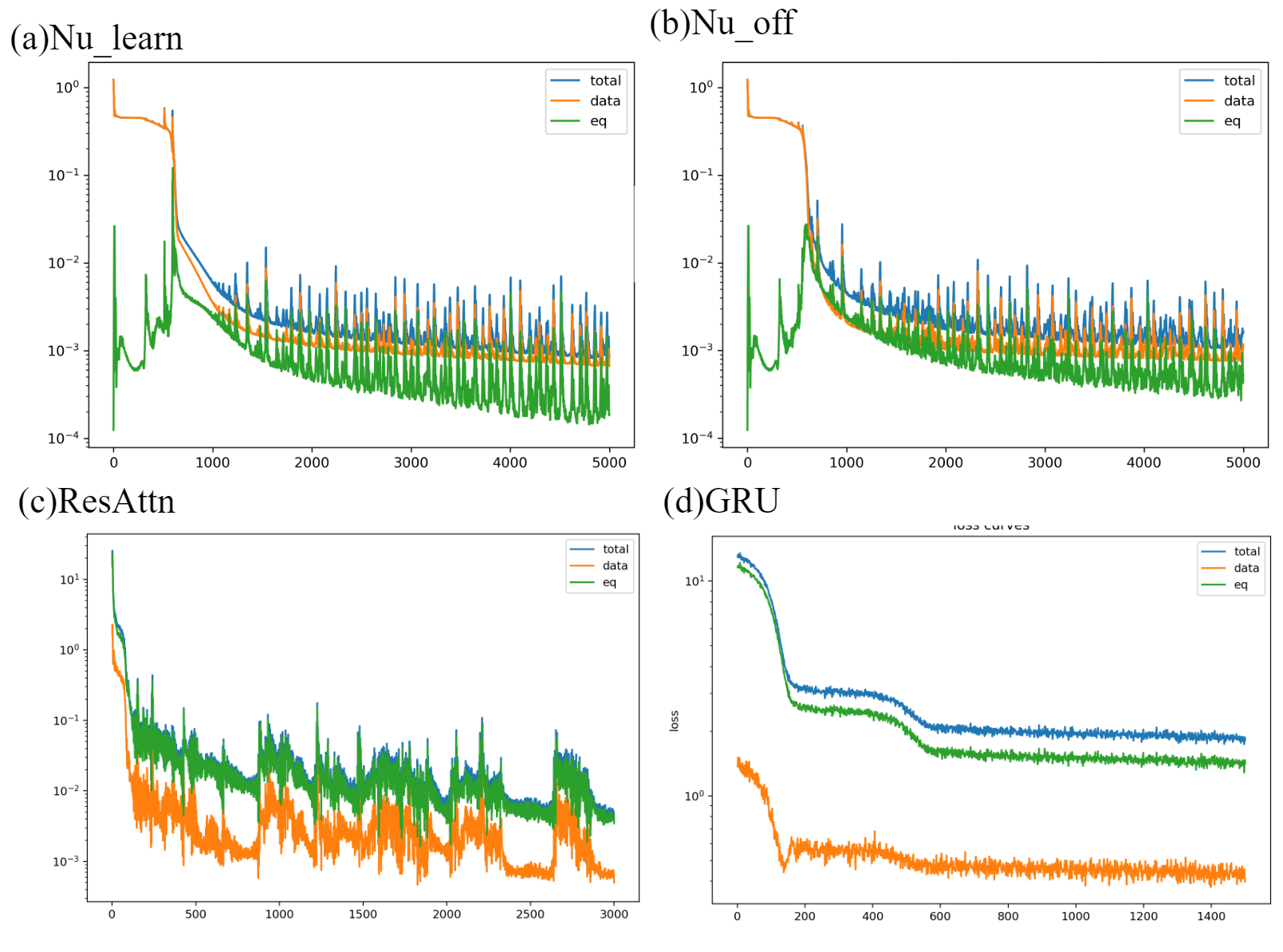}
    \caption{Training loss histories for the Kovasznay flow under the four compared methods: (a) Nu\_learn, (b) Nu\_off, (c) ResAttn, and (d) GRU. For each method, the total loss, data loss, and equation loss are shown.}
    \label{fig:kovasznay_loss}
\end{figure}

\begin{figure}[H]
    \centering
    \includegraphics[width=0.80\columnwidth]{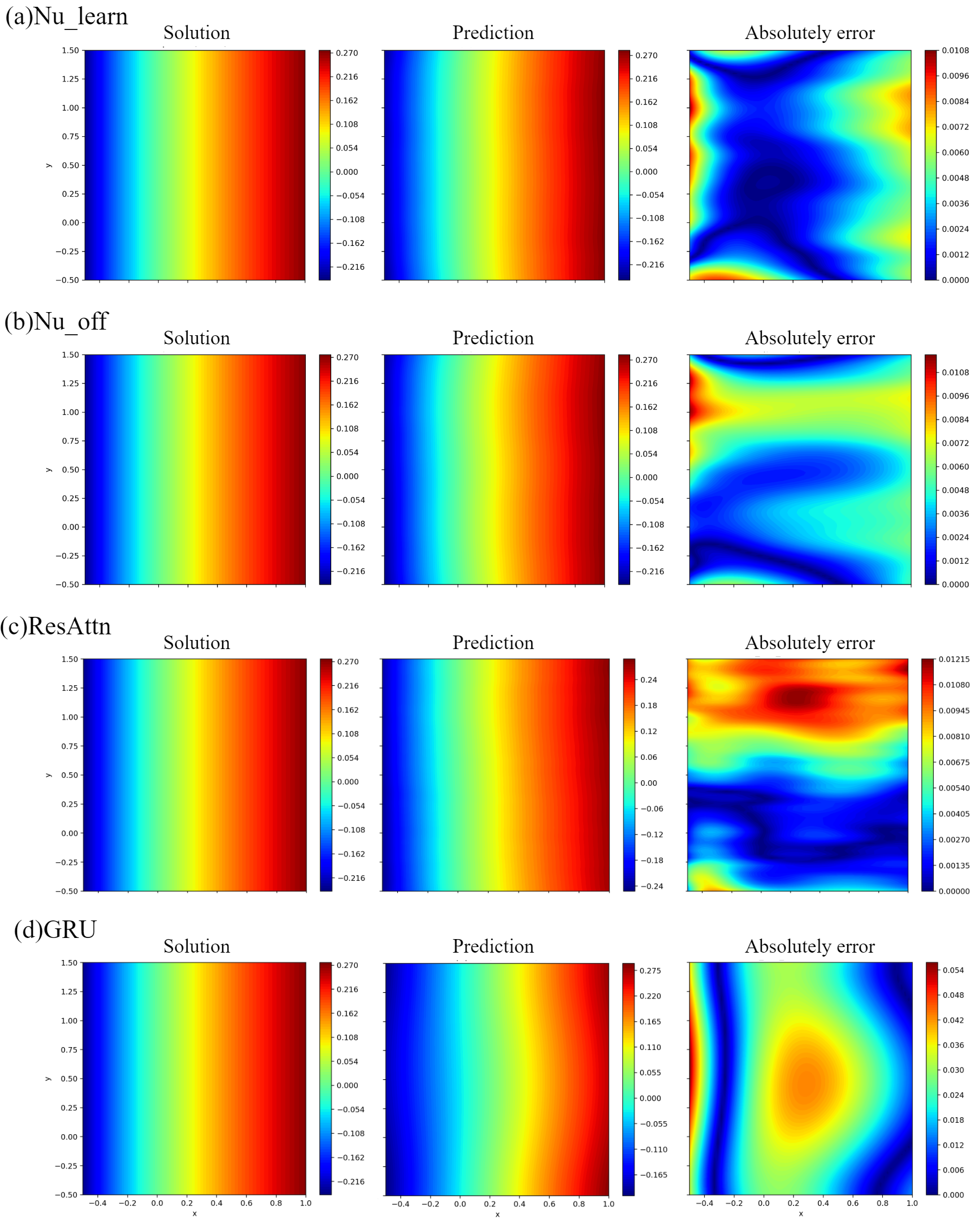}
    \caption{Pressure-field reconstruction for the Kovasznay flow. In each row, from left to right: reference solution, prediction, and absolute error. The four rows correspond to (a) Nu\_learn, (b) Nu\_off, (c) ResAttn, and (d) GRU, respectively.}
    \label{fig:kovasznay_p}
\end{figure}

\begin{figure}[H]
    \centering
    \includegraphics[width=0.80\columnwidth]{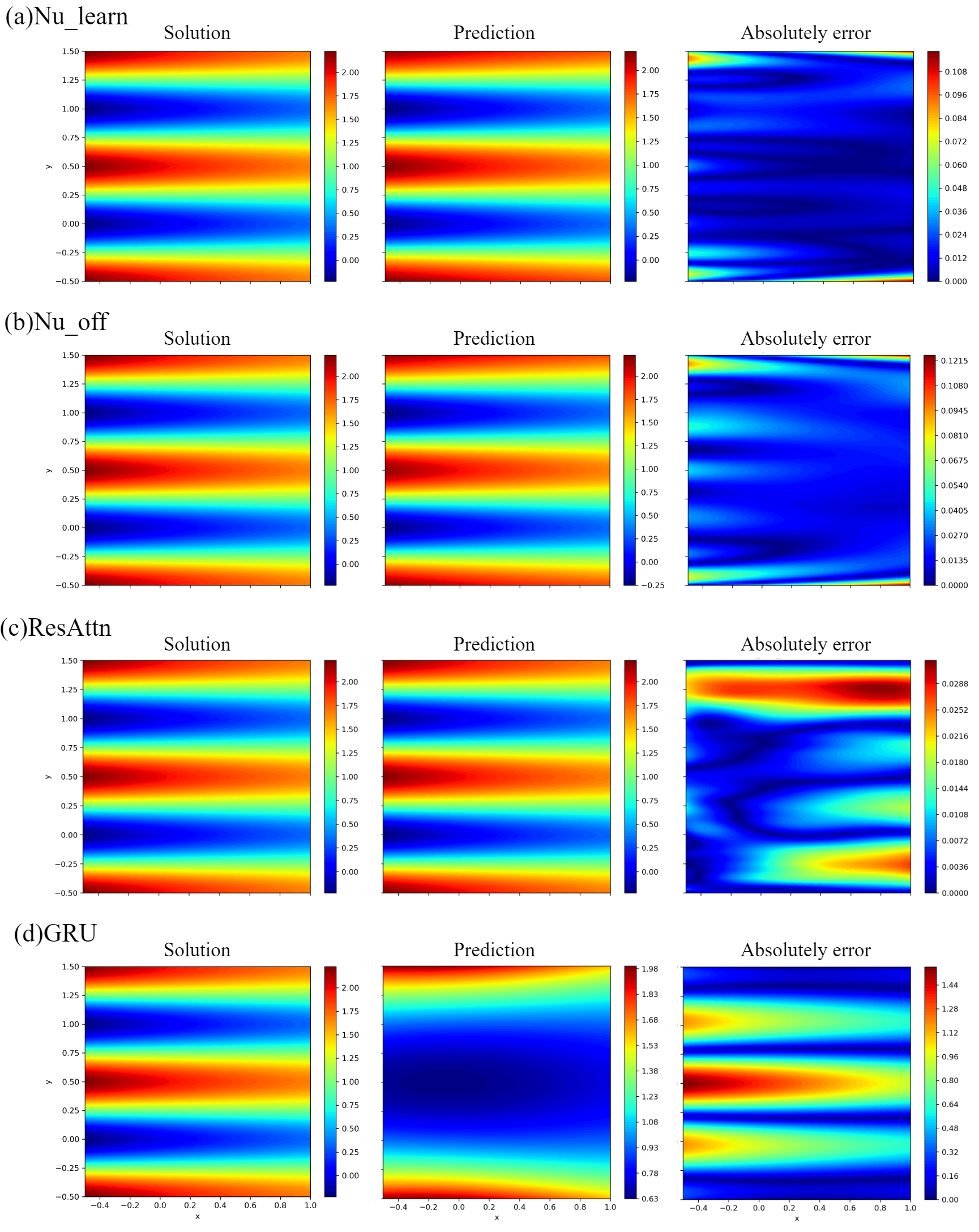}
    \caption{Reconstruction of the streamwise velocity component $u$ for the Kovasznay flow. In each row, from left to right: reference solution, prediction, and absolute error. The four rows correspond to (a) Nu\_learn, (b) Nu\_off, (c) ResAttn, and (d) GRU, respectively.}
    \label{fig:kovasznay_u}
\end{figure}

\begin{figure}[H]
    \centering
    \includegraphics[width=0.80\columnwidth]{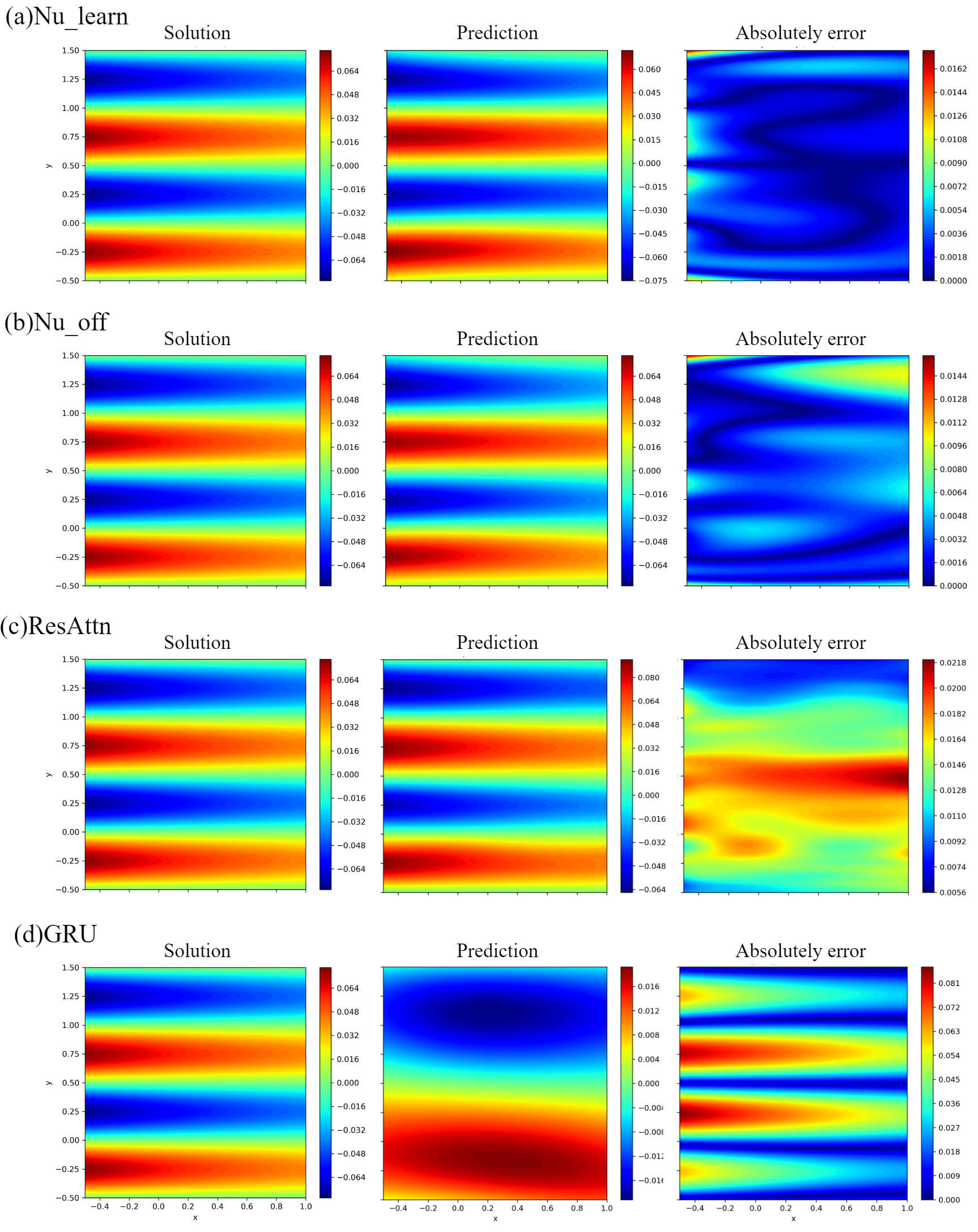}
    \caption{Reconstruction of the transverse velocity component $v$ for the Kovasznay flow. In each row, from left to right: reference solution, prediction, and absolute error. The four rows correspond to (a) Nu\_learn, (b) Nu\_off, (c) ResAttn, and (d) GRU, respectively.}
    \label{fig:kovasznay_v}
\end{figure}

For the unforced Kovasznay flow, while all four computational methods successfully recover the global flow topology, substantial discrepancies emerge regarding optimization stability and ultimate reconstruction accuracy. As illustrated in Fig.~\ref{fig:kovasznay_loss}, both Nu\_learn and Nu\_off exhibit highly stable convergence behaviors, with Nu\_learn driving the equation residual to the lowest absolute level during the late stages of training. In stark contrast, the ResAttn model demonstrates noticeably stronger oscillations, and the GRU architecture stagnates at comparatively high total loss and equation loss levels throughout the entire optimization process. The reconstructed pressure and velocity fields, visualized in Figs.~\ref{fig:kovasznay_p} through \ref{fig:kovasznay_v}, corroborate these training dynamics. Among the evaluated methods, Nu\_learn produces the cleanest spatial error distributions across all three physical variables, with the Nu\_off configuration consistently ranking as the second most accurate. Conversely, the error regions associated with ResAttn are markedly more diffuse, and the GRU baseline delivers the weakest overall reconstruction quality, particularly within the detailed velocity fields. The quantitative error metrics reported in Table~\ref{tab:kovasznay} are fully consistent with these visual observations, confirming that for the Kovasznay benchmark, the active modulation model not only outperforms its own ablation setting but also demonstrates clear superiority over both alternative backbone architectures.

\begin{table}[H]
\centering
\caption{Prediction errors for the Kovasznay flow.}
\label{tab:kovasznay}
\setlength{\tabcolsep}{4pt}
\small
\begin{tabular}{lccc}
\toprule
Model & $E_{L2}(u)$ & $E_{L2}(v)$ & $E_{L2}(p)$ \\
\midrule
Nu\_learn & 1.582771 & 0.226508 & 0.302667 \\
Nu\_off   & 2.099818 & 0.299187 & 0.366258 \\
ResAttn   & 0.831740 & 0.944840 & 0.400800 \\
GRU       & 41.680820 & 2.586000 & 1.680830 \\
\bottomrule
\end{tabular}
\end{table}

\begin{figure}[H]
    \centering
    \includegraphics[width=0.80\columnwidth]{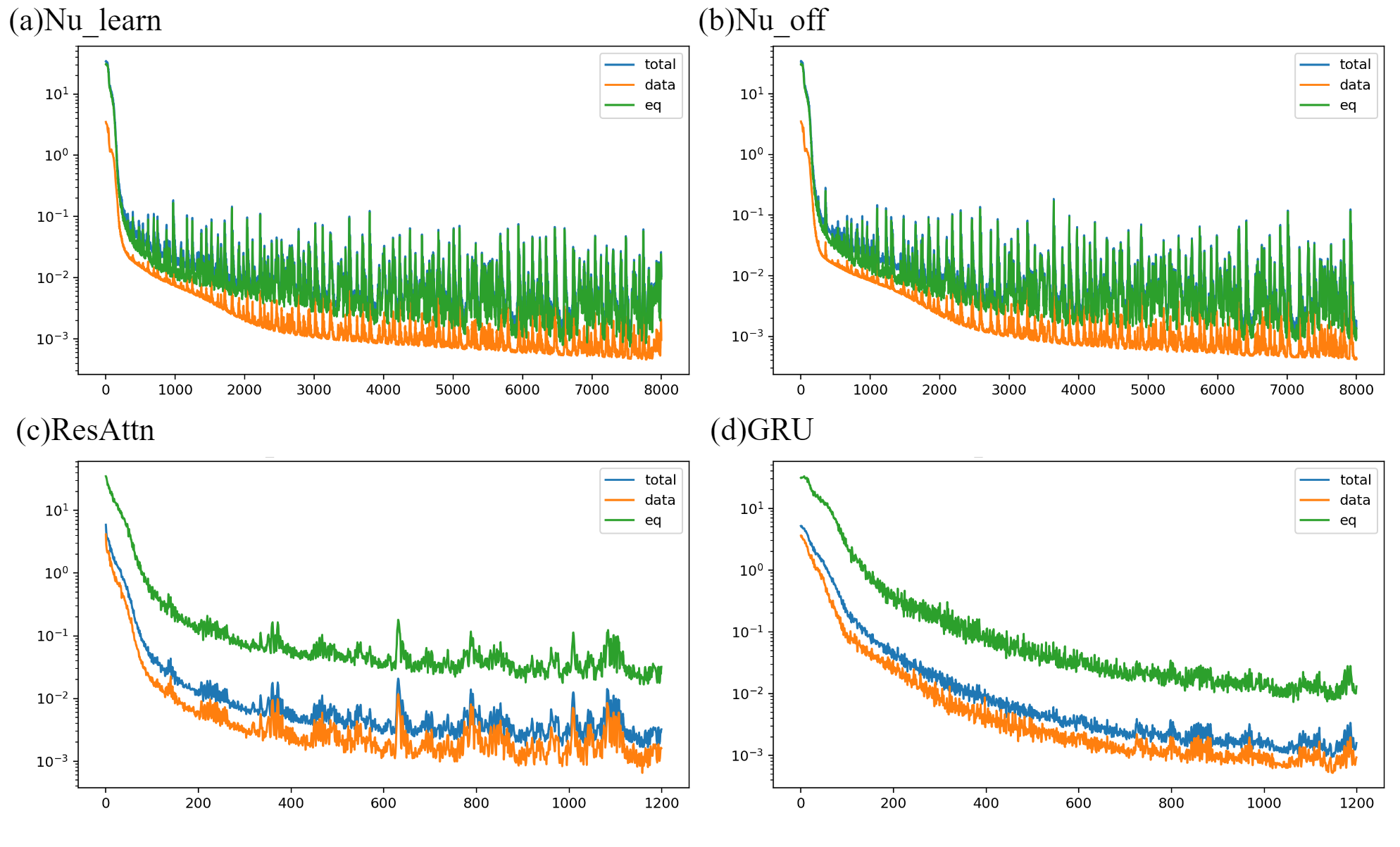}
    \caption{Training loss histories for manufactured forcing flow I under the four compared methods: (a) Nu\_learn, (b) Nu\_off, (c) ResAttn, and (d) GRU. For each method, the total loss, data loss, and equation loss are shown.}
    \label{fig:f1_loss}
\end{figure}

\begin{figure}[H]
    \centering
    \includegraphics[width=0.80\columnwidth]{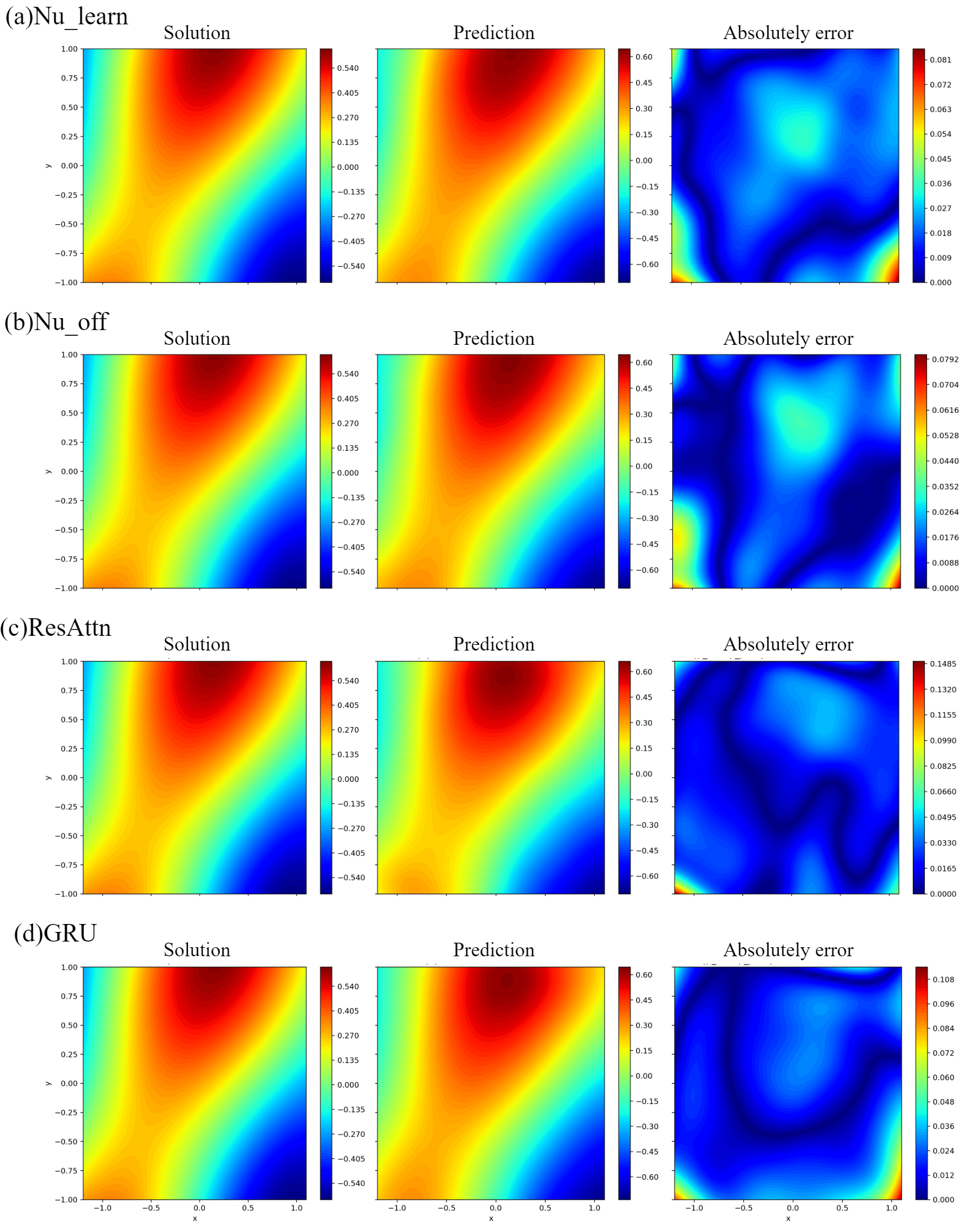}
    \caption{Pressure-field reconstruction for manufactured forcing flow I. In each row, from left to right: reference solution, prediction, and absolute error. The four rows correspond to (a) Nu\_learn, (b) Nu\_off, (c) ResAttn, and (d) GRU, respectively.}
    \label{fig:f1_p}
\end{figure}

\begin{figure}[H]
    \centering
    \includegraphics[width=0.80\columnwidth]{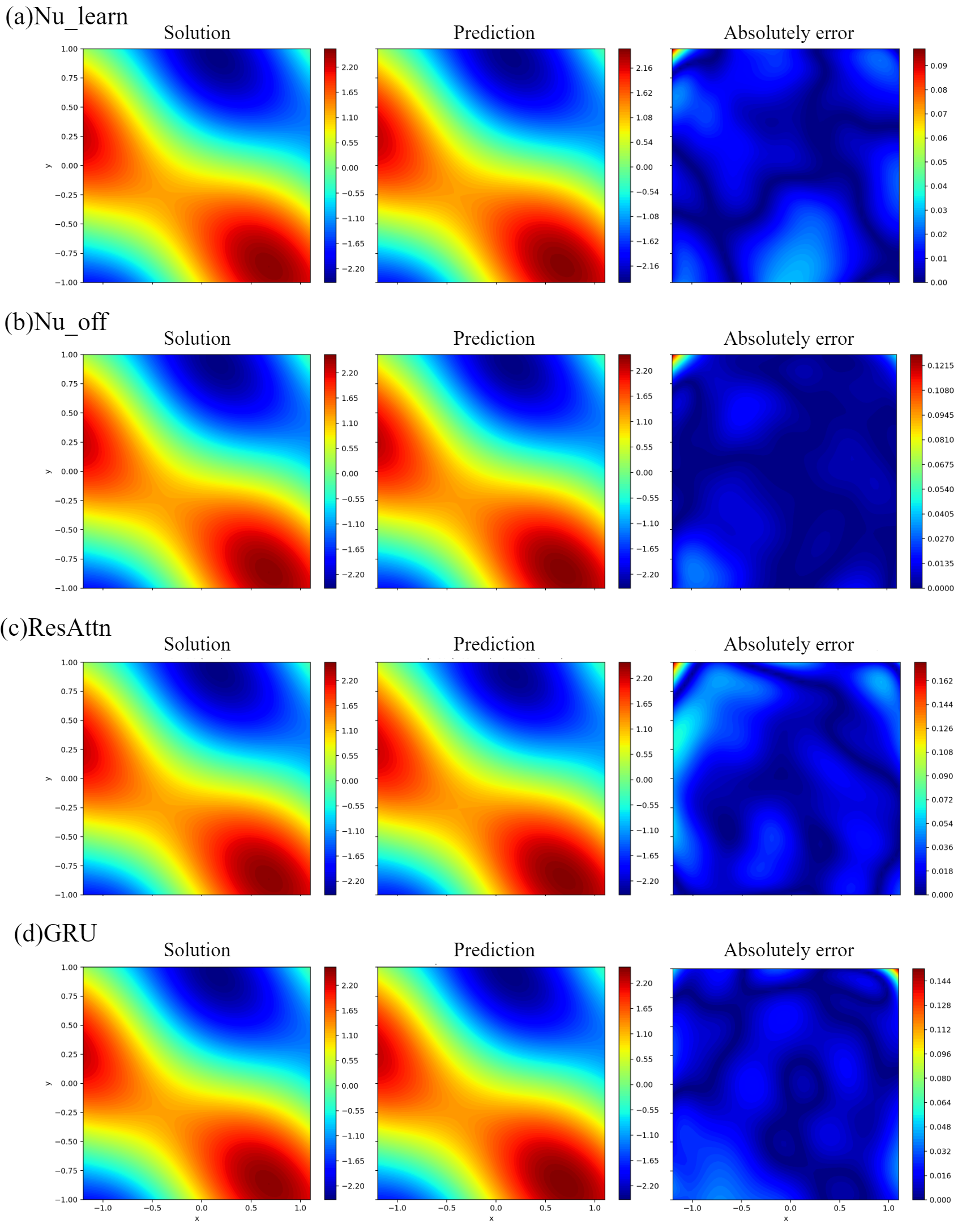}
    \caption{Reconstruction of the streamwise velocity component $u$ for manufactured forcing flow I. In each row, from left to right: reference solution, prediction, and absolute error. The four rows correspond to (a) Nu\_learn, (b) Nu\_off, (c) ResAttn, and (d) GRU, respectively.}
    \label{fig:f1_u}
\end{figure}

\begin{figure}[H]
    \centering
    \includegraphics[width=0.80\columnwidth]{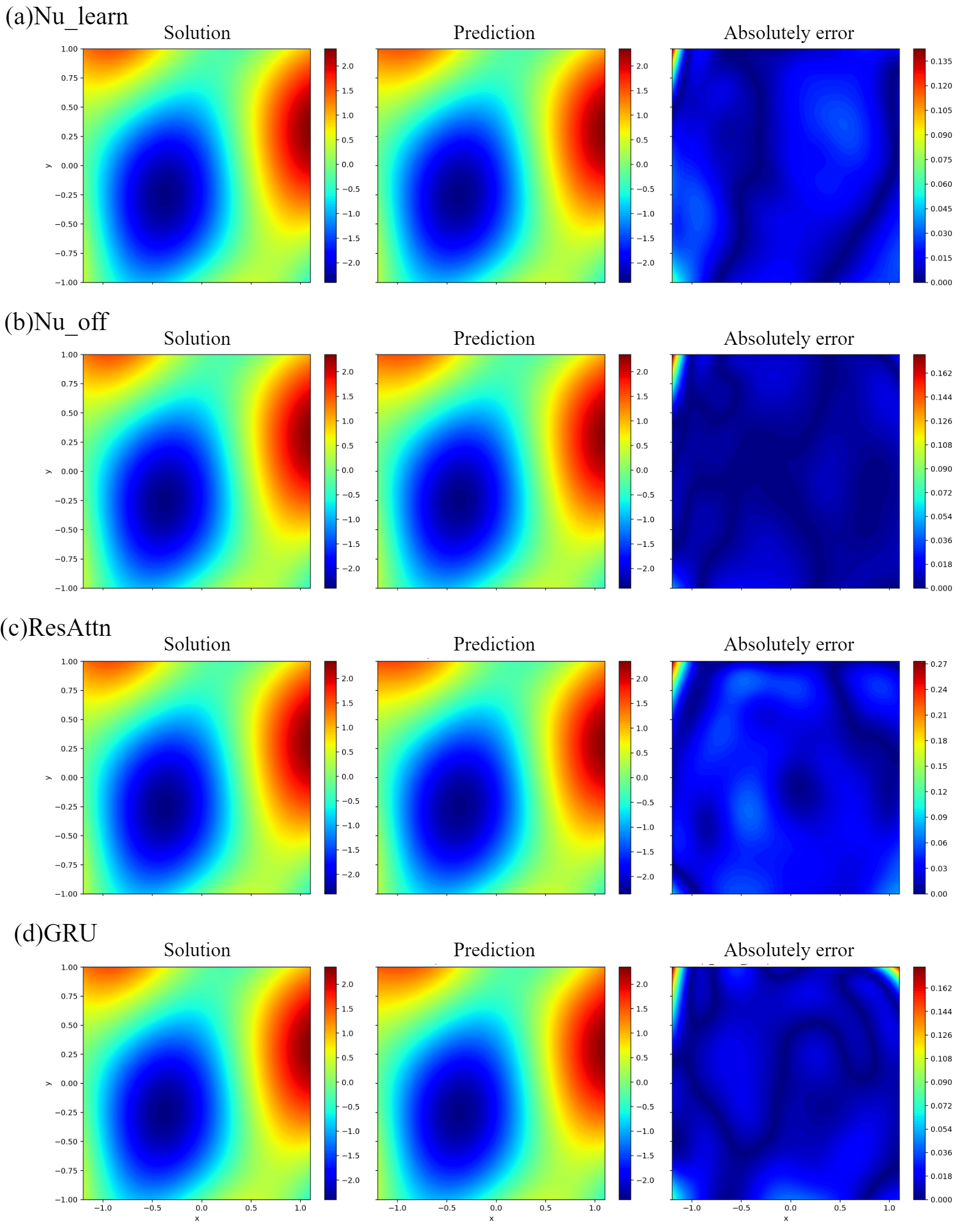}
    \caption{Reconstruction of the transverse velocity component $v$ for manufactured forcing flow I. In each row, from left to right: reference solution, prediction, and absolute error. The four rows correspond to (a) Nu\_learn, (b) Nu\_off, (c) ResAttn, and (d) GRU, respectively.}
    \label{fig:f1_v}
\end{figure}

Transitioning to manufactured forcing flow I, all models manage to capture the dominant spatial structure of the targeted analytical solution, yet the variations in predictive accuracy remain pronounced. According to Fig.~\ref{fig:f1_loss}, Nu\_learn and Nu\_off again maintain the most orderly and stable training processes, with significantly reduced oscillatory behavior relative to the ResAttn and GRU baselines. In particular, Nu\_learn achieves the smallest terminal equation loss, revealing the strongest capacity to preserve physical consistency under the more demanding forced-flow setting. This optimization advantage is mirrored in the reconstructed field visualizations shown in Figs.~\ref{fig:f1_p} through \ref{fig:f1_v}. The Nu-based models generate comparatively compact and localized error distributions, whereas both baseline architectures exhibit broader error spreading and visibly less faithful field recovery. The quantitative data summarized in Table~\ref{tab:f1} further demonstrate that Nu\_learn attains the most favorable overall balance across the three physical variables, thereby confirming that the proposed modulation mechanism remains highly effective when the governing equations are augmented by nontrivial forcing terms.

\begin{table}[H]
\centering
\caption{Prediction errors for manufactured forcing flow I.}
\label{tab:f1}
\setlength{\tabcolsep}{4pt}
\small
\begin{tabular}{lccc}
\toprule
Model & $E_{L2}(u)$ & $E_{L2}(v)$ & $E_{L2}(p)$ \\
\midrule
Nu\_learn & 0.706361 & 1.426821 & 0.803229 \\
Nu\_off   & 0.919142 & 1.675384 & 0.826457 \\
ResAttn   & 1.474770 & 2.119820 & 1.443130 \\
GRU       & 0.938620 & 1.093330 & 1.350770 \\
\bottomrule
\end{tabular}
\end{table}

\begin{figure}[H]
    \centering
    \includegraphics[width=0.80\columnwidth]{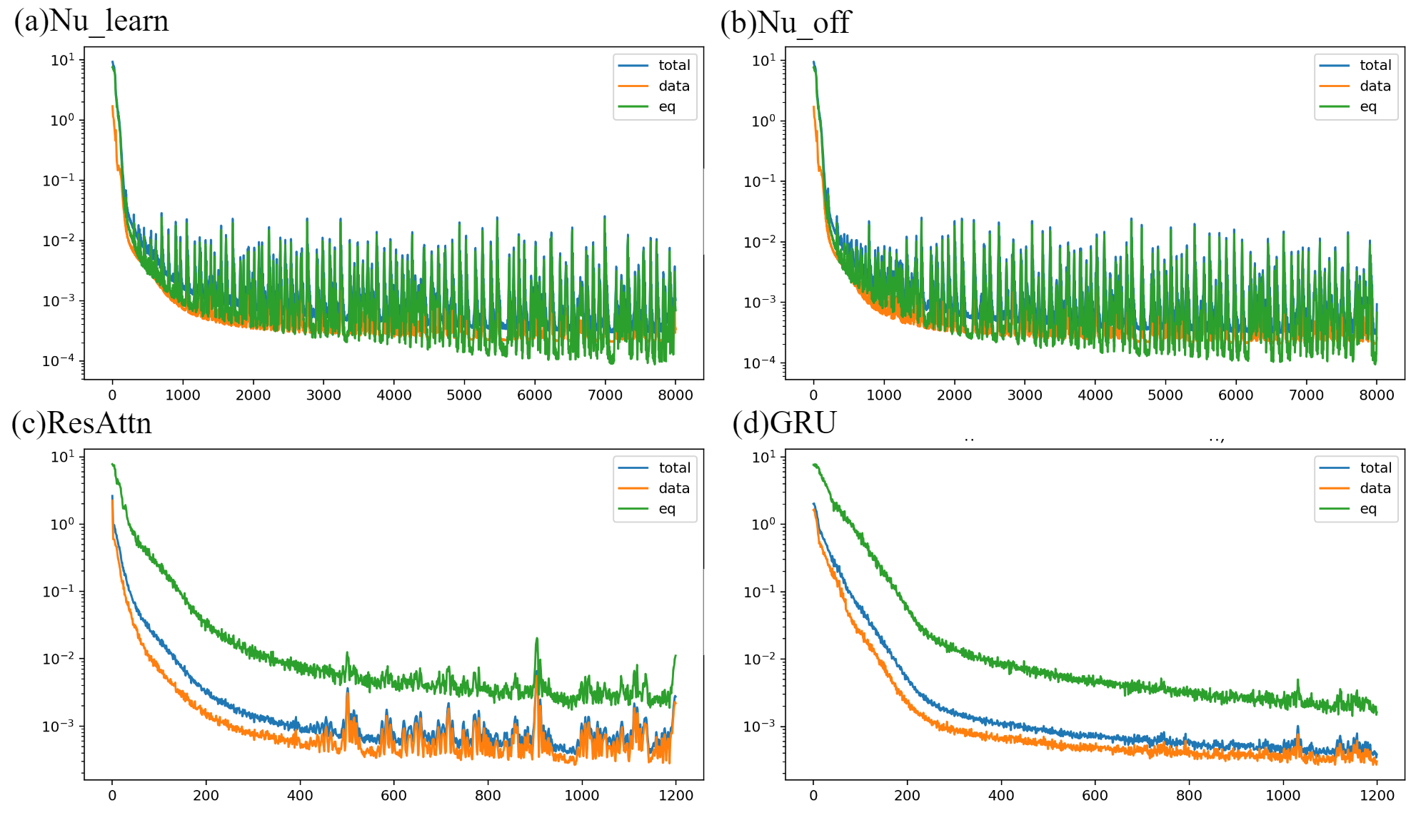}
    \caption{Training loss histories for manufactured forcing flow II under the four compared methods: (a) Nu\_learn, (b) Nu\_off, (c) ResAttn, and (d) GRU. For each method, the total loss, data loss, and equation loss are shown.}
    \label{fig:f2_loss}
\end{figure}

\begin{figure}[H]
    \centering
    \includegraphics[width=0.80\columnwidth]{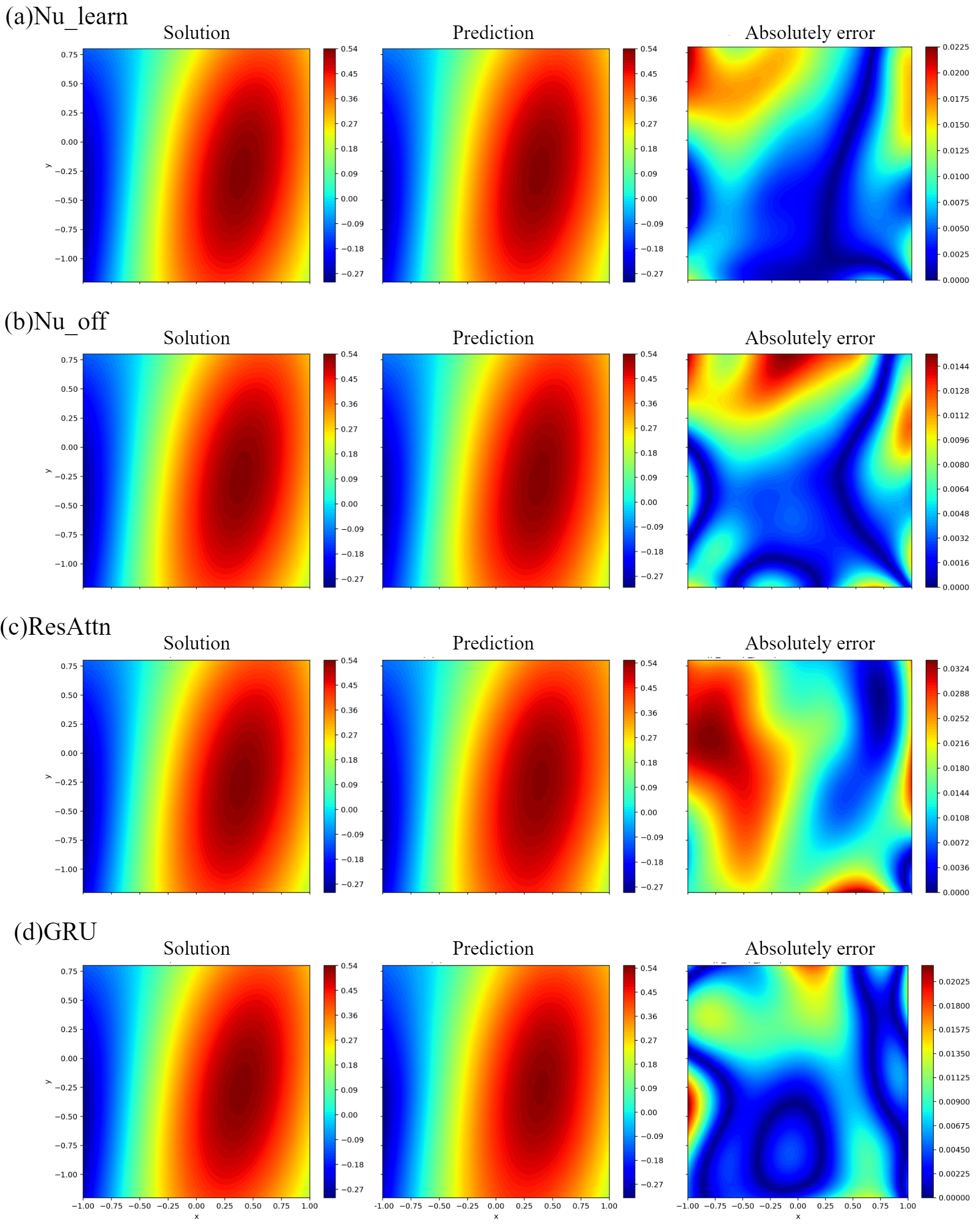}
    \caption{Pressure-field reconstruction for manufactured forcing flow II. In each row, from left to right: reference solution, prediction, and absolute error. The four rows correspond to (a) Nu\_learn, (b) Nu\_off, (c) ResAttn, and (d) GRU, respectively.}
    \label{fig:f2_p}
\end{figure}

\begin{figure}[H]
    \centering
    \includegraphics[width=0.80\columnwidth]{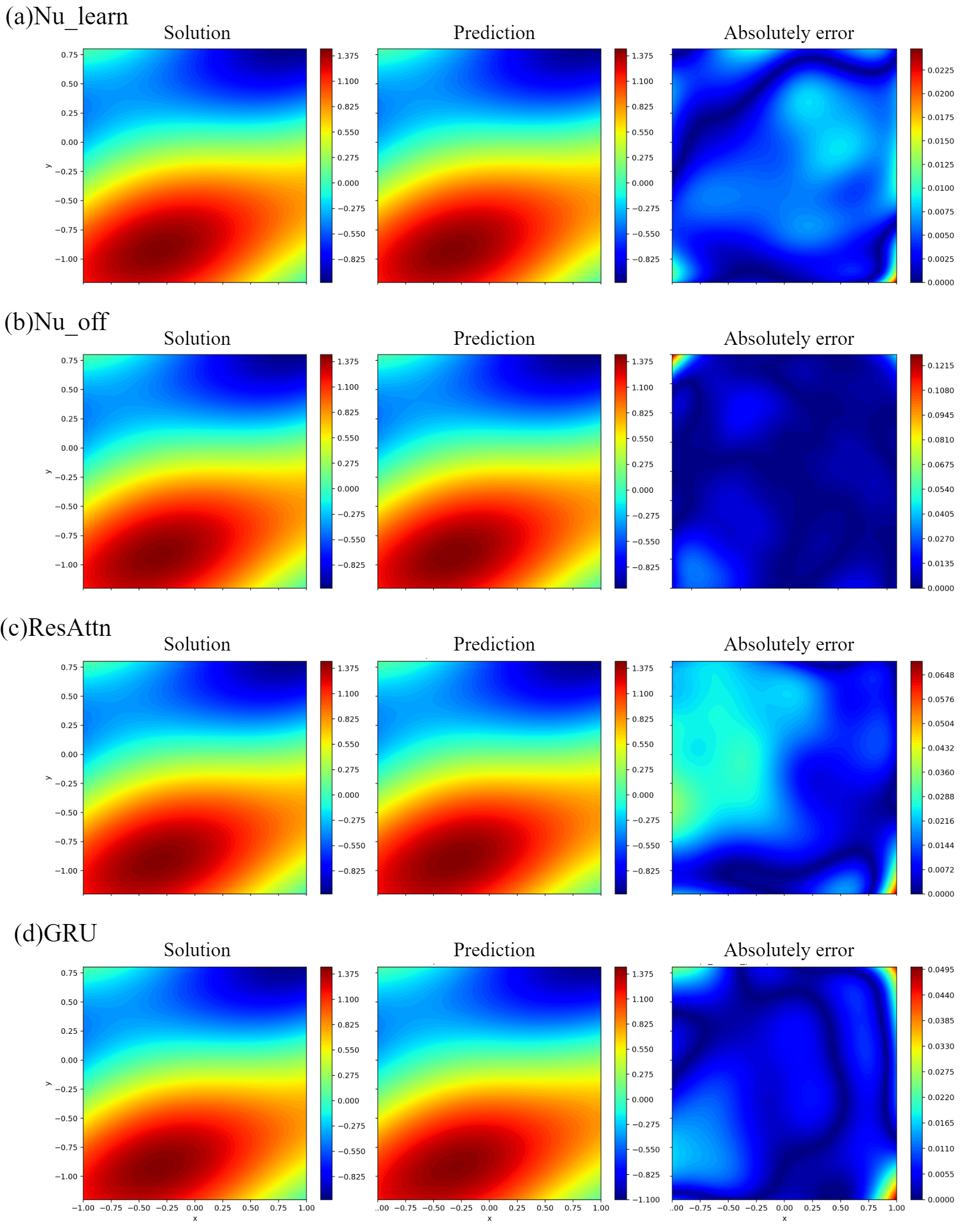}
    \caption{Reconstruction of the streamwise velocity component $u$ for manufactured forcing flow II. In each row, from left to right: reference solution, prediction, and absolute error. The four rows correspond to (a) Nu\_learn, (b) Nu\_off, (c) ResAttn, and (d) GRU, respectively.}
    \label{fig:f2_u}
\end{figure}

\begin{figure}[H]
    \centering
    \includegraphics[width=0.90\columnwidth]{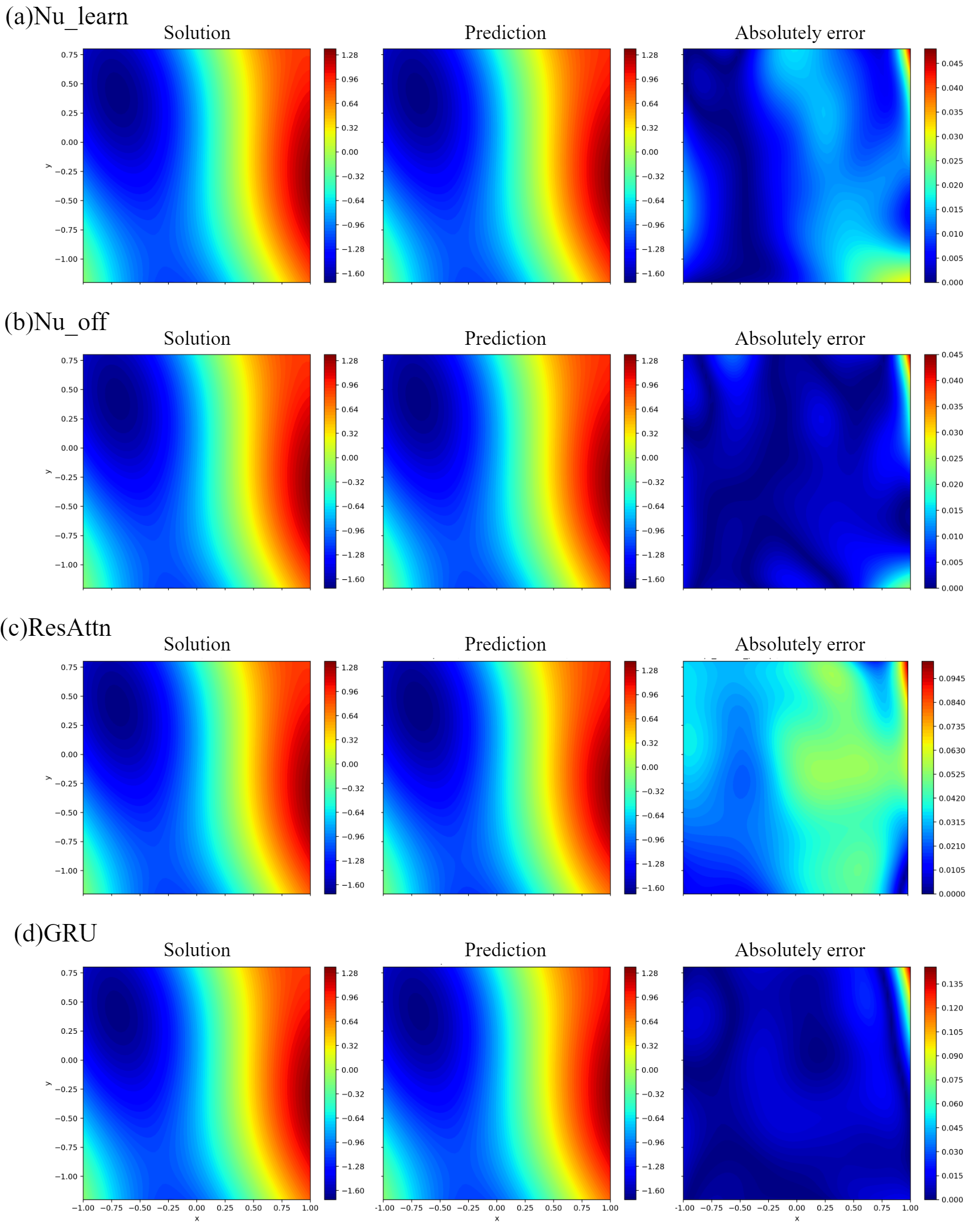}
    \caption{Reconstruction of the transverse velocity component $v$ for manufactured forcing flow II. In each row, from left to right: reference solution, prediction, and absolute error. The four rows correspond to (a) Nu\_learn, (b) Nu\_off, (c) ResAttn, and (d) GRU, respectively.}
    \label{fig:f2_v}
\end{figure}

In the context of manufactured forcing flow II, all evaluated methods successfully resolve the large-scale spatial distribution of the target flow, although the performance gap narrows slightly compared to the preceding cases due to the smoother inherent dynamics. As depicted in Fig.~\ref{fig:f2_loss}, Nu\_learn and Nu\_off persist as the two most stable training models, achieving both total and equation losses that are distinctively lower than those produced by ResAttn and GRU. The corresponding reconstructed fields in Figs.~\ref{fig:f2_p} through \ref{fig:f2_v} visually confirm that the modulation-based models continue to outpace the baseline backbones by maintaining tighter error distributions. In contrast, ResAttn and GRU generate comparatively diffuse error topologies characterized by broader areas of medium to high discrepancy. This overarching pattern is directly reflected in the quantitative error norms detailed in Table~\ref{tab:f2}, proving that even under smoother forced flow conditions, the Nu-based framework maintains superior optimization characteristics and stronger multi-field reconstruction capabilities than conventional structural alternatives.

\begin{table}[H]
\centering
\caption{Prediction errors for manufactured forcing flow II.}
\label{tab:f2}
\setlength{\tabcolsep}{4pt}
\small
\begin{tabular}{lccc}
\toprule
Model & $E_{L2}(u)$ & $E_{L2}(v)$ & $E_{L2}(p)$ \\
\midrule
Nu\_learn & 0.140874 & 0.346145 & 0.176153 \\
Nu\_off   & 0.101997 & 0.371105 & 0.169394 \\
ResAttn   & 1.064360 & 2.331090 & 1.215460 \\
GRU       & 0.471710 & 0.816000 & 0.428160 \\
\bottomrule
\end{tabular}
\end{table}

Synthesizing the results across all three distinct benchmarks reveals a robust and consistent pattern of performance enhancement. The proposed LVM-PINN framework, Nu\_learn and Nu\_off, are systematically superior to both the ResAttn and GRU architectures, clearly indicating that the proposed LVM-PINN framework delivers heightened optimization stability and refined reconstruction quality regardless of the specific flow regime. Moreover, the active Nu\_learn configuration demonstrates the clearest performance advantage on both the complex Kovasznay flow and manufactured forcing flow I, whereas its relative improvement over the ablation setting becomes slightly more variable-dependent in the context of manufactured forcing flow II. Nevertheless, the combined empirical evidence derived from the convergence histories, qualitative field visualizations, and rigorous quantitative metrics consistently validates the efficacy of the proposed modulation framework for solving diverse incompressible flow problems.

\section{Conclusion}

In this study, we developed an advanced physics-informed neural network framework that integrates a learnable viscosity modulation mechanism to reconstruct incompressible flow fields from sparse and noisy observational data. By dynamically embedding an auxiliary spatiotemporal scalar field directly into the viscous diffusion term of the momentum equations, the proposed model adaptively regulates local dissipation while strictly preserving the exact mathematical structure of the Navier-Stokes equations. Comprehensive numerical evaluations across the classical Kovasznay flow and two manufactured forcing benchmarks demonstrated that this modulated approach consistently delivers superior optimization stability and higher fidelity flow reconstructions compared to established residual-attention and recurrent neural network baselines. Furthermore, rigorous controlled ablation studies revealed that the active modulation configuration provides a distinct advantage in capturing complex nonlinear flow structures, particularly in unforced regimes, though its relative improvement can become more variable-dependent in certain smoother forced dynamics. Ultimately, these findings validate that incorporating adaptive physical mechanisms into the residual formulation fundamentally enhances the robustness and accuracy of neural network solvers for complex fluid dynamics applications.

\section*{Declaration of Competing Interest}
The authors declare that they have no known competing financial interests or personal relationships that could have appeared to influence the work reported in this paper.

\section*{Acknowledgments}
This work was supported by the Science and Technology Development Project of Jilin Province (20250102032JC).

\section*{Data Availability}
The code used in this study is openly available at the following GitHub repository:
\url{https://github.com/Uderwood-TZ/LVM-PINN.git}

\appendix
\section{Analytical Expressions of the Manufactured Forcing Cases}
\label{app:manufactured_cases}

For reproducibility, this appendix provides the explicit analytical expressions used in the manufactured forcing benchmarks. In both cases, the analytical velocity, pressure, and reference modulation fields are prescribed first, and the forcing term is then constructed by substituting these fields into the momentum residual operator. More specifically, the forcing components are generated according to
\begin{equation}
\begin{aligned}
f_x ={}& u_t + u u_x + v u_y + p_x \\
&- \frac{1}{Re}(1+Nu^\ast)(u_{xx}+u_{yy}),
\end{aligned}
\label{eq:fx_appendix}
\end{equation}
\begin{equation}
\begin{aligned}
f_y ={}& v_t + u v_x + v v_y + p_y \\
&- \frac{1}{Re}(1+Nu^\ast)(v_{xx}+v_{yy}),
\end{aligned}
\label{eq:fy_appendix}
\end{equation}
so that the prescribed analytical fields satisfy the governing equations exactly.

\paragraph{Manufactured forcing flow I.}
The computational domain is
\begin{equation}
x \in [-1.2,1.1], \qquad y \in [-1.0,1.0], \qquad t \in [0,1],
\end{equation}
with Reynolds number
\begin{equation}
Re = 2500.
\end{equation}
The streamfunction is prescribed as
\begin{equation}
\begin{aligned}
\psi(x,y,t) ={}&
0.90\sin(1.20x+2.10y+0.70t+0.30) \\
&+ 0.55\cos(2.40x-1.10y-1.15t+0.20) \\
&+ 0.16\sin(0.60xy+0.45t+0.40),
\end{aligned}
\end{equation}
and the incompressible velocity field is obtained from
\begin{equation}
u = \frac{\partial \psi}{\partial y},
\qquad
v = -\frac{\partial \psi}{\partial x}.
\end{equation}
Equivalently, the explicit expressions are
\begin{equation}
\begin{aligned}
u(x,y,t) ={}&
1.89\cos(1.20x+2.10y+0.70t+0.30) \\
&- 0.605\sin(2.40x-1.10y-1.15t+0.20) \\
&+ 0.096x\cos(0.60xy+0.45t+0.40),
\end{aligned}
\end{equation}
\begin{equation}
\begin{aligned}
v(x,y,t) ={}&
-1.08\cos(1.20x+2.10y+0.70t+0.30) \\
&- 1.32\sin(2.40x-1.10y-1.15t+0.20) \\
&- 0.096y\cos(0.60xy+0.45t+0.40).
\end{aligned}
\end{equation}
The pressure field is defined as
\begin{equation}
\begin{aligned}
p(x,y,t) ={}&
0.45\cos(1.70x-0.80y+0.55t+0.10) \\
&+ 0.18\sin(0.70x+1.60y-0.90t+0.60) \\
&+ 0.04xy + 0.03x,
\end{aligned}
\end{equation}
and the reference modulation field is given by
\begin{equation}
\begin{aligned}
Nu^\ast(x,y,t) ={}&
0.22 \\
&+ 0.08\sin(0.90x-1.30y+0.60t+0.10) \\
&+ 0.05\cos(1.60x+0.70y-0.50t+0.40).
\end{aligned}
\end{equation}
The forcing components $(f_x,f_y)$ are then obtained from Eqs.~\eqref{eq:fx_appendix}--\eqref{eq:fy_appendix}.

\paragraph{Manufactured forcing flow II.}
The computational domain is
\begin{equation}
x \in [-1.0,1.0], \qquad y \in [-1.2,0.8], \qquad t \in [0,1],
\end{equation}
with Reynolds number
\begin{equation}
Re = 2800.
\end{equation}
The analytical velocity field is prescribed as
\begin{equation}
\begin{aligned}
u(x,y,t) ={}&
0.36\cos(1.8x+0.5y+0.95t+0.1) \\
&- 0.902\cos(0.7x-2.2y+0.55t+0.8) \\
&- 0.06x\sin(0.5xy-0.6t+0.2) + 0.18,
\end{aligned}
\end{equation}
\begin{equation}
\begin{aligned}
v(x,y,t) ={}&
-1.296\cos(1.8x+0.5y+0.95t+0.1) \\
&- 0.287\cos(0.7x-2.2y+0.55t+0.8) \\
&+ 0.06y\sin(0.5xy-0.6t+0.2) - 0.11,
\end{aligned}
\end{equation}
The pressure field is defined as
\begin{equation}
\begin{aligned}
p(x,y,t) ={}&
0.28\cos(1.3x+0.6y-0.7t+0.3) \\
&+ 0.25\sin(2.0x-0.9y+0.4t+0.2) \\
&+ 0.015x^2 - 0.02y,
\end{aligned}
\end{equation}
and the reference modulation field is prescribed by
\begin{equation}
\begin{aligned}
Nu^\ast(x,y,t) ={}&
0.24 \\
&+ 0.05\sin(0.60x+1.40y+0.25t+0.50) \\
&+ 0.05\cos(1.90x-0.80y+0.55t+0.40).
\end{aligned}
\end{equation}
As in manufactured forcing flow I, the forcing components are generated from Eqs.~\eqref{eq:fx_appendix}--\eqref{eq:fy_appendix}, ensuring that the prescribed analytical fields satisfy the governing equations exactly.

\bibliographystyle{aipnum4-1}
\bibliography{aipsamp}
\end{document}